\documentclass[aps,prb,twocolumn,showpacs,superscriptaddress]{revtex4}
\usepackage{graphicx}
\usepackage{dcolumn}
\usepackage{bm}
\usepackage{amssymb}
\usepackage{amsmath}

\begin{document}

\title{The upper critical field and its anisotropy in RbCr$_{3}$As$_{3}$}
\author{Qimei Liang}
\affiliation{Anhui Province Key Laboratory of Condensed Matter Physics at Extreme Conditions, High Magnetic Field Laboratory of the Chinese Academy of Sciences, Hefei 230031,
China}
\affiliation{University of Science and Technology of China, Hefei 230026, China}
\author{Tong Liu}
\affiliation{Beijing National Laboratory for Condensed Matter Physics, Institute of
Physics, Chinese Academy of Sciences, Beijing 100190, China}
\affiliation{School of Physical Sciences, University of Chinese Academy of Sciences,
Beijing 100049, China}
\author{Chuanying Xi}
\affiliation{Anhui Province Key Laboratory of Condensed Matter Physics at Extreme Conditions, High Magnetic Field Laboratory of the Chinese Academy of Sciences, Hefei 230031,
China}
\author{Yuyan Han}
\affiliation{Anhui Province Key Laboratory of Condensed Matter Physics at Extreme Conditions, High Magnetic Field Laboratory of the Chinese Academy of Sciences, Hefei 230031,
China}
\author{Gang Mu}
\affiliation{State Key Laboratory of Functional Materials for Informatics, Shanghai Institute of Microsystem and Information Technology, Chinese Academy of Sciences, Shanghai 200050, China}
\affiliation{CAS Center for Excellence in Superconducting Electronics(CENSE), Shanghai 200050, China}
\author{Li Pi}
\affiliation{Anhui Province Key Laboratory of Condensed Matter Physics at Extreme Conditions, High Magnetic Field Laboratory of the Chinese Academy of Sciences, Hefei 230031,
China}
\author{Zhi-An Ren}
\email{renzhian@iphy.ac.cn}
\affiliation{Beijing National Laboratory for Condensed Matter Physics, Institute of
Physics, Chinese Academy of Sciences, Beijing 100190, China}
\affiliation{School of Physical Sciences, University of Chinese Academy of Sciences,
Beijing 100049, China}
\author{Zhaosheng Wang}
\email{zswang@hmfl.ac.cn}
\affiliation{Anhui Province Key Laboratory of Condensed Matter Physics at Extreme Conditions, High Magnetic Field Laboratory of the Chinese Academy of Sciences, Hefei 230031,
China}

\begin{abstract}
The temperature dependence of the upper critical field ($H_{c2}$) in RbCr$%
_{3}$As$_{3}$ single crystals ($T_{c}\approx $ 7.3 K) has been determined by
means of magnetoresistance measurements with temperature down to 0.35 K in
static magnetic fields up to 38 T. The magnetic field was applied both for directions parallel ($H\parallel c
$, $H_{c2}^{\parallel c}$) and perpendicular ($H\perp c$, $H_{c2}^{\perp c}$) to the
Cr chains. The curves $H_{c2}^{\parallel c}(T)$ and $H_{c2}^{\perp c}(T)$ cross at $\sim $ 5.5 K. As a result, the anisotropy parameter $\gamma
(T)=H_{c2}^{\perp c}/H_{c2}^{\parallel c}(T)$ increases from 0.5 near $T_{c}$ to 1.6
at low temperature. Fitting with the Werthamer-Helfand-Hohenberg (WHH) model
yields zero-temperature critical fields of $\mu_0H_{c2}^{\parallel c}(0)\approx $ 27.2 T
and $\mu_0H_{c2}^{\perp c}(0)\approx $ 43.4 T, both exceeding the BCS weak-coupling
Pauli limit $\mu_0H_{p}=1.84T_{c}=13.4$ T.  The results indicate that the paramagnetic pair breaking effect is strong for $H \parallel c$ but absent for $H \perp c$, which was further confirmed by the angle dependent magnetoresistance and $H_{c2}$ measurements.

%All of the results are similar to K$_2$Cr$_3$As$_3$,

%The angle-dependent resistance shows the maxmium moves from
%along the $c\ $direction to perpendicular to the $c\ $direction with
%increasing the magnetic field, indicating an unconventional supercoductivity
%in this system.
\end{abstract}

\maketitle

\section{Introduction}

Recently, superconductivity was found in Cr-based ternary compounds A$_2$Cr$_3$As$_3$ at ambient pressure\cite{Cao-1,Cao-2,Cao-3,Ren-3} following the discovery of superconductivity in CrAs at a critical pressure $P_{c}$ $\approx$ 8 kbar.\cite{CrAs} A$_2$Cr$_3$As$_3$ compounds have a quasi-one-dimensional (Q1D) hexagonal noncentrosymmetric crystal structure with a space group of $P\overline{6}m2$. The infinite [(Cr$_3$As$_3$)$^{2-}$]$_{\infty}$ linear chains are separated by alkali-metal cations. For A = Na, K, Rb, and Cs, the superconducting $T_{c}$ is 8.6 K, 6.1 K, 4.8 K and 2.2 K, respectively.\cite{Cao-1,Cao-2,Cao-3,Ren-3} As showing very particular crystal structure and unconventional superconducting properties, this Cr-based superconducting family has attracted intense interests.\cite{Multiband,Molecular,NMR,Nodal-K2Cr3As3,3band-K2Cr3As3,K2Cr3As3_canfield,K2Cr3As3_canfield1, K2Cr3As3_zhu, K2Cr3As3_cao, Stongcoupling-K2Cr3As3, p-wave, usr, Raman} However, the experimental results within the context of pairing symmetry have not yet reached a consensus\cite{review,NLWang}. A$_2$Cr$_3$As$_3$ superconductors are extremely reactive when exposed in air, probably due to the existence of ¡°crowded¡± A1 atoms in the crystal structure.\cite{Cao-1} The samples are easily oxidized during most experimental procedures, which hinders many further studies for their intrinsic physical characteristics.

Lately, by deintercalating half of the A$^{+}$ ions using ethanol from the A$_2$Cr$_3$As$_3$ lattice, another type of Q1D compounds ACr$_3$As$_3$ (A = K, Rb, Cs) with similar crystal structure were obtained, with $T_{c} \approx$ 5 K and 7.3 K for KCr$_{3}$As$_{3}$ and RbCr$_{3}$As$_{3}$ .\cite{Ren-1,Ren-2} Unlike the A$_2$Cr$_3$As$_3$ compounds, ACr$_3$As$_3$ superconductors have a centrosymmetric lattice with the space group $P6_3/m$ and are air-stable.\cite{NLWang,Ren-1,Ren-2} Recent neutron and x-ray diffraction measurements show that the superconductivity in KCr$_3$As$_3$ is induced by Hydrogen doping.\cite{KHCe3As3} Density functional theory (DFT) analysis
shows that KH$_x$Cr$_3$As$_3$ has a similar electronic structure to K$_2$Cr$_3$As$_3$.\cite{KHCe3As3}
Thus it is important to study the superconducting properties of ACr$_3$As$_3$ and compare to the A$_2$Cr$_3$As$_3$ compounds.  As a basic
parameter, the temperature dependence of the upper critical field, $H_{c2}$,
reflects the underlying electronic structure responsible for
superconductivity and provides valuable information on the microscopic
origin of pair breaking. By measuring the temperature dependence of $%
H_{c2}$ of RbCr$_{3}$As$_{3}$, information on the superconducting pairing mechanism of ACr$_3$As$_3$ superconductors
can be gained.

In this work, we present temperature and magnetic field dependent magnetoresistance measurements with magnetic
fields applied parallel and perpendicular to the $c$ axis, and angle dependent magnetoresistance measurements on RbCr$_{3}$As$_{3}$
single crystals. $H_{c2}$ was determined over a wide range of temperatures down to 0.35 K in static magnetic fields up
to 38 T. We find that the curves $H_{c2}^{\parallel c}(T)$ and $H_{c2}^{\perp c}(T)$ cross at $\sim $ 5.5 K, and both $H_{c2}^{\parallel c}(0)$ and $H_{c2}^{\perp c}(0)$ exceed the BCS weak-coupling Pauli limit. The results indicate that the paramagnetic pair breaking effect is strong for $H \parallel c$ but absent for $H \perp c$, which was further confirmed by the angle dependent magnetoresistance and $H_{c2}$ measurements.

\section{Experiment}

Single crystals of RbCr$_{3}$As$_{3}$ were prepared by the deintercalation of Rb$^+$ ions from Rb$_{2}$Cr$_{3}$As$_{3}$ precursors, which were grown out of the RbAs and CrAs mixture using a high temperature solution growth method.\cite{Canfield-1} The asgrown Rb$_{2}$Cr$_{3}$As$_{3}$ single crystals were immersed in pure dehydrated ethanol and kept for one week for the fully
deintercalation of Rb$^+$ ions at room temperature. The
obtained samples were washed by ethanol thoroughly. To
further improve the sample quality, the as-prepared crystals
were annealed in an evacuated quartz tube at 373K for
10 h.\cite{Ren-1} All the experimental procedures were performed
in a glove box filled with high-purity Ar gas to avoid introducing
impurities. More detailed information can be found in Ref.22.
The obtained RbCr$_{3}$As$_{3}$ crystals are needle-like with a typical size of $5 \times 0.2 \times 0.18$ mm$^3$, and
quite stable in air at room temperature.

\begin{figure}[tbp]
\includegraphics[width=0.85\columnwidth]{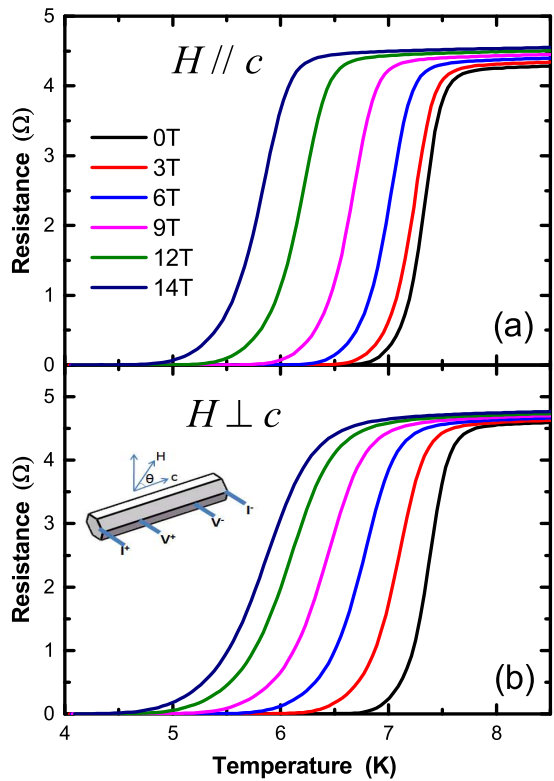}
\caption{Temperature dependence of resistance for RbCr$_3$As$_3$ single crystal A1 at fields
 $\mu_{0}H$ = 0, 3, 6, 9, 12, 14 T with (a) $H\parallel c$ and (b) $H\perp c$, respectively. The inset of (b) illustrates the definition of angle $\theta$.}
\label{fig:fig1}
\end{figure}

The resistance was measured by a standard four-probe method with a current $I = 100$ $\mu$A flowing along the $c$ axis(as shown in the inset of Fig.\ \ref{fig:fig1} (b)).  Magnetic fields were applied parallel and perpendicular to the $c$ axis($H\parallel c$, $H\parallel I$ and $H\perp c$, $H\perp I$).
The temperature and angular dependence of resistance was measured by use of a commercial Physical Property
Measurement System (PPMS) with magnetic fields up to 14 T. In the angle dependent measurements, $\theta = 0^o$ corresponded to the configuration of $H \parallel c$ axis and $\theta = 90^o$ to $H\perp c$ axis, respectively. The field dependent resistance measurements shown in Fig. \ \ref{fig:fig2}
were carried out at temperatures down to 0.35 K with a $^3$He cryostat in High Magnetic Field Laboratory of Chinese Academy of Science.
A water-cooling magnet which generates the maximum magnetic field up to
38.5 T was employed. The samples were fixed on the sample holder with GE-7031 varnish. A delta mode system with Keithley models 6221 and 2182A was used.

\section{Results and Discussion}

\begin{figure}[tbp]
\includegraphics[width=0.9\columnwidth]{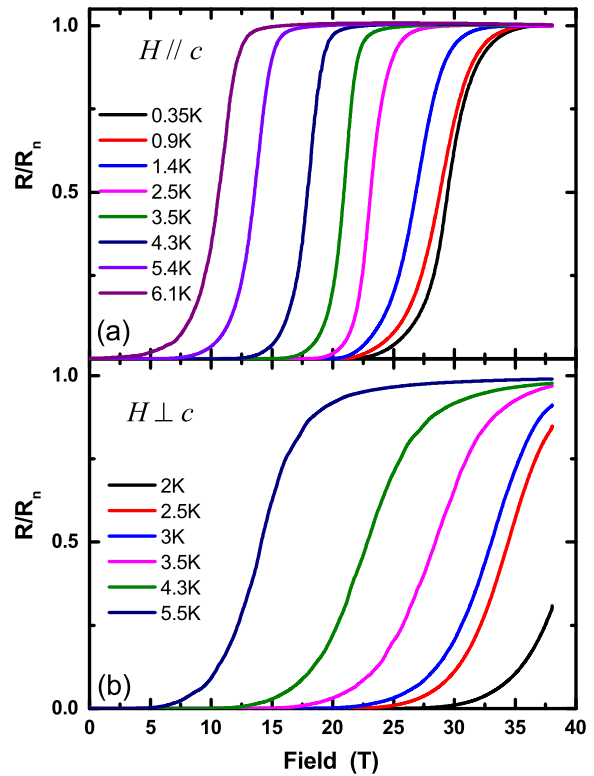}
\caption{Magnetic field dependence of resistance for RbCr$_3$As$_3$ single crystal A1 at different temperatures with (a) $H \parallel c$ and (b) $H\perp c$ up to 38 T. The data are normalized to
the value in the normal state $R_n$ (the resistance at 6.1 K and 38 T).}
\label{fig:fig2}
\end{figure}

We measured five RbCr$_{3}$As$_{3}$ samples from two batches (labeled as A1, A2, A3 and B1, B2).  All samples show similar behaviors.
Fig.\ \ref{fig:fig1} shows a typical result of the temperature dependent resistance in magnetic fields from 0 to 14 T for $H \parallel c $ and $H \perp c$, respectively. The magnetic field shifts the
zero-resistance state to lower temperature a bit slower for $H \parallel c$
than for $H \perp c$ at temperature close to $T_c$. As temperature decreasing, the case is reversed, which implies a reversal of the anisotropy of $H_{c2}$.

The magnetic field dependent resistances measured at different
temperatures in static magnetic fields up to 38 T for $H \parallel c$ and $H \perp c$ are shown in Fig.\ \ref{fig:fig2}(a) and Fig.\ \ref{fig:fig2}
(b), respectively. The sample is the same one shown in Fig.\ \ref{fig:fig1}. Apparently, 38 T is enough to suppress superconductivity completely at temperature down to 0.35 K for $H \parallel c$. However, a stronger field
is needed to suppress superconductivity for $H \perp c$ at low temperatures. Thus a reversal of the anisotropy of $H_{c2}$ has been confirmed. As the current flowed along the $c$ axis during the measurements, it was Lorentz force free for $H \parallel c$. However, for $H \perp c$, there was a maximum of Lorentz force, which could generate a motion of the vortices and lead to a finite resistance region.\cite{vortex}  This region is called the vortex-liquid phase, which broadens the resistive transitions.\cite{vortex liquid}
As shown in Figs.\ \ref{fig:fig1} and\ \ref{fig:fig2}, field-induced broadenings of the resistive
transitions are small, suggesting a very narrow vortex-liquid region in RbCr$_3$As$_3$. This behavior is similar to A$_{2}$Cr$_{3}$As$_{3}$ compounds\cite{K2Cr3As3_canfield,K2Cr3As3_canfield1, K2Cr3As3_zhu,K2Cr3As3_cao} and some Fe-based superconductors like Ba122,\cite{YuanHQ, KanoM, WangBaKFeAs, WangBaFeNiAs} FeTe$_{0.6}$Se$_{0.4}$,\cite{Khim} and LiFeAs.\cite{Khim2} In order to reduce the influence of
the vortex-liquid phase and superconducting fluctuations, the temperature or field where the normal-state resistance $R_n$ is reduced to 50\,\% was chosen as the criterion to determine the $H_{c2}-T$ phase diagram.

\begin{figure}[tbp]
\includegraphics[width=0.9\columnwidth]{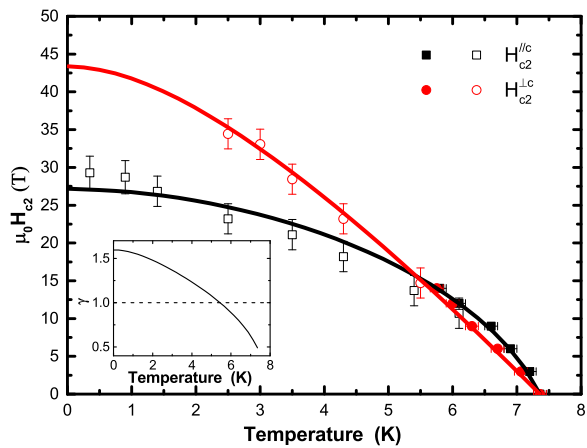}
\caption{Temperature dependence of $H_{c2}$ extracted from the magnetoresistance measurements for RbCr$_3$As$_3$ single crystal A1. The solid symbols are obtained from PPMS measurements, and the open symbols are obtained from water-cooling magnet measurements. The black and red solid line show  WHH fits for $H_{c2}^{\parallel c}$ and $H_{c2}^{\perp c}$ with fitting parameters $\alpha=8$,
$\lambda_{so}=1.6$ and $\alpha=0$, $\lambda_{so}=0$, respectively. The inset shows the anisotropy parameter $\gamma(T)$ calculated from the fitting results.}
\label{fig:fig3}
\end{figure}

The resulting critical fields $H_{c2}^{\parallel c}$ and $H_{c2}^{\perp c}$ are
summarized in Fig.\ \ref{fig:fig3}. The closed symbols are obtained from PPMS measurements by use of temperature scans, and the open symbols are obtained from water-cooling magnet measurements utilizing magnetic field scans. The slopes $\mu_0H^{'} = \mu_0 dH^{c}_{c2}/dT_c$ at $T_c$ are -18 T/K and -8.5 T/K for $H \parallel c$ and $H \perp c$ , respectively. According to the Ginzburg-Landau(GL) theory, the effective mass anisotropy $m_{\perp}/m_{\parallel} = (H_{c2}^{\parallel c'}/H_{c2}^{\perp c'})^2 \approx 4.5$. This anisotropy value is only about one-sixth of that in Rb$_{2}$Cr$_{3}$As$_{3}$ \cite{K2Cr3As3_cao}, revealing an reduced Q1D character in RbCr$_{3}$As$_{3}$. According to the
Werthamer-Helfand-Hohenberg (WHH) formula,\cite{WHH} $H_{c2}^{orb} = -0.73T_c(dH_{c2}/dT)$$\mid_{T_c}$. Using the GL relations, $H_{c2}^{orb,\parallel c}(0) = \Phi_0/(2\pi\xi_{\perp c}^{2})$ and $H_{c2}^{orb,\perp c}(0) = \Phi_0/(2\pi\xi_{\perp c}\xi_{\parallel c})$, where $\Phi_0$ is the magnetic flux quantum, the anisotropic coherence lengths can be estimated as $\xi_{\perp c}(0) \approx 1.9$ nm and $\xi_{\parallel c}(0) \approx 3.9$ nm, respectively. These values are close to the results reported in Rb$_{2}$Cr$_{3}$As$_{3}$ \cite{K2Cr3As3_cao}. The $\xi_{\perp c}(0)$ value is about twice of the
interchain distance,\cite{Ren-2} indicating a uniaxially anisotropic 3D superconductivity.
As temperature decreases, the curves $H_{c2}^{\parallel c}(T)$ and $H_{c2}^{\perp c}(T)$ cross at $T\approx $ 5.5 K. For a weak coupling conventional BCS superconductor, the Pauli-limiting field can be estimated by \cite{Clogston} $\mu_0H_{p} = 1.84 T_{c}$ T, resulting in $\mu_0H_{p}$ = 13.4 T.  From Fig.\ \ref{fig:fig3} one can see, $H_{c2}^{\parallel c}(0)$ and $H_{c2}^{\perp c}(0)$ are larger than the Pauli-limiting by two and three times respectively. These results are similar to the results of A$_2$Cr$_3$As$_3$,\cite{K2Cr3As3_canfield,K2Cr3As3_canfield1,K2Cr3As3_zhu,K2Cr3As3_cao} indicating comparable strong electron correlation in the Cr-based family.

\begin{figure}[tbp]
\includegraphics[width=0.9\columnwidth]{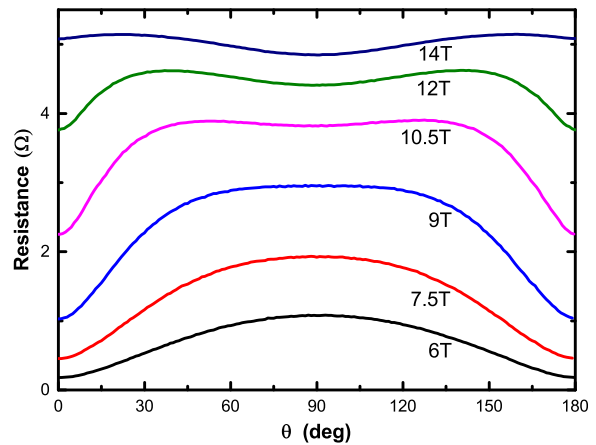}
\caption{Angular dependence of resistance at 6.5 K with  magnetic field $\mu_0H$ = 6, 7.5, 9, 10.5, 12 and 14 T for RbCr$_3$As$_3$ single crystal A2.}
\label{fig:fig4}
\end{figure}

To quantitatively describe our results, we use the full WHH formula
that incorporates the spin-paramagnetic effect via the Maki parameter
$\alpha $ and the spin-orbit scattering constant $\lambda_{so}$ to describe the
experimental $H_{c2}(T)$ data:\cite{WHH} %\begin{equation}
\begin{multline}
\ln\frac{1}{t} = \sum_{\nu=-\infty}^{\infty} \{ \frac{1}{|2\nu+1|}- \\
\left[|2\nu+1| + \frac{\bar{h}}{t} + \frac{(\alpha\bar{h}/t)^{2}}{|2\nu+1|+(%
\bar{h}+\lambda_{so})/t}\right]^{-1} \}  \label{eq:whh}
\end{multline}
where $t=T/T_c$ and $\bar{h}=(4/\pi^{2})[H_{c2}/|dH_{c2}/dT|_{T_c}]$. As shown by the solid line in Fig.\ \ref{fig:fig3}, the best fit ($\alpha=8$,
$\lambda_{so}=1.6$ and $\alpha=0$, $\lambda_{so}=0$) can reproduce the experimental data
well, resulting in $\mu_{0}H_{c2}^{\parallel c}(0)$ = 27.2 T and $\mu_{0}H_{c2}^{\perp c}(0)$ = 43.4 T, respectively. The results indicate that Pauli pair breaking is strong for $H_{c2}^{\parallel c}(T)$ but absent for $H_{c2}^{\perp c}(T)$. As $\alpha \propto \gamma_n \rho_n$,\cite{WHH} where $\gamma_n$ and $\rho_n$ are the normal state electronic specific heat coefficient and the normal state dc resistivity respectively, the large $\alpha$ is consistent with the high $\gamma_n$ and $\rho_n$ reported in the ACr$_{3}$As$_{3}$ compounds.\cite{Ren-1, LiSY} According to the Maki formula,\cite{Maki} for $H \parallel c$, $\alpha = \sqrt{2}H_{c2}^{orb,\parallel c}(0)/H_{p}$ = 10. The fitting result is a bit smaller than the value calculated from the Maki formula. This deviation has been widely observed in Fe-based superconductors, and been considered to be a consequence of the enhancement of $H_p$ over $H_p^{BCS}$ due to the strong coupling effect.\cite{Mugang}

The anisotropy parameter $\gamma
(T)=H_{c2}^{\perp c}/H_{c2}^{\parallel c}(T)$ can be calculated from the fitting results. $\gamma$ increases from 0.5 near $T_{c}$ to $> 1$ below $T\approx $ 5.5 K where the $H_{c2}(T)$ curves cross, and about 1.6 at low temperature.  Similar behaviors of $\gamma$ have
been reported in A$_2$Cr$_3$As$_3$,\cite{K2Cr3As3_canfield,K2Cr3As3_zhu,K2Cr3As3_cao} heavy-fermion superconductor UPt$_3$,\cite{UPt3} and Q1D superconductors Li$_{0.9}$Mo$_6$O$_7$ \cite{LiMoO} and organic superconductors $(TMTSF)_2PF_6$. \cite{(TMTSF)2PF6} Recently, DFT calculations find strong structural instabilities of KCr$_{3}$As$_{3}$, which would lead to a much more one-dimensional Fermi surface structure.\cite{DFT} However, comparing to $\gamma$ = 0.19 near $T_{c}$ in Rb$_{2}$Cr$_{3}$As$_{3}$ ,\cite{K2Cr3As3_cao} $\gamma$ = 0.5 clearly indicates weaker anisotropy
in RbCr$_{3}$As$_{3}$ which possesses smaller interchain distance.

According to the anisotropic Ginzburg-Landau
theory, the effective-mass anisotropy leads to the anisotropy of the orbital limited upper critical field $H_{c2}^{GL}(\theta) = H_{c2}^{\parallel c}/\sqrt{cos^{2}(\theta)+\gamma^{-2}sin^{2}(\theta)}$, and the resistivity in the mixed state depends on the effective
field $H/H_{c2}^{GL}(\theta)$.\cite{Blatter} Thus the maximum and minimum of the angle dependent resistance should be at $\theta = 0^{o}$ or 90$^{o}$ depending on $\gamma$ $> 1$ or $< 1$.  Fig.\ \ref{fig:fig4} presents angle dependent resistance at 6.5 K for RbCr$_3$As$_3$ single crystal A2. When the magnetic field is less than 9 T, the maximum of the resistance is at $\theta = 90^{o}$.  However, a hollow shows up at $\theta = 90^{o}$ as magnetic field increasing further, indicating that there is a strong anisotropic paramagnetic pair-breaking effect in this system.  These results are consistent with the results shown in Fig.\ \ref{fig:fig3}.

\begin{figure}[tbp]
\includegraphics[width=0.85\columnwidth]{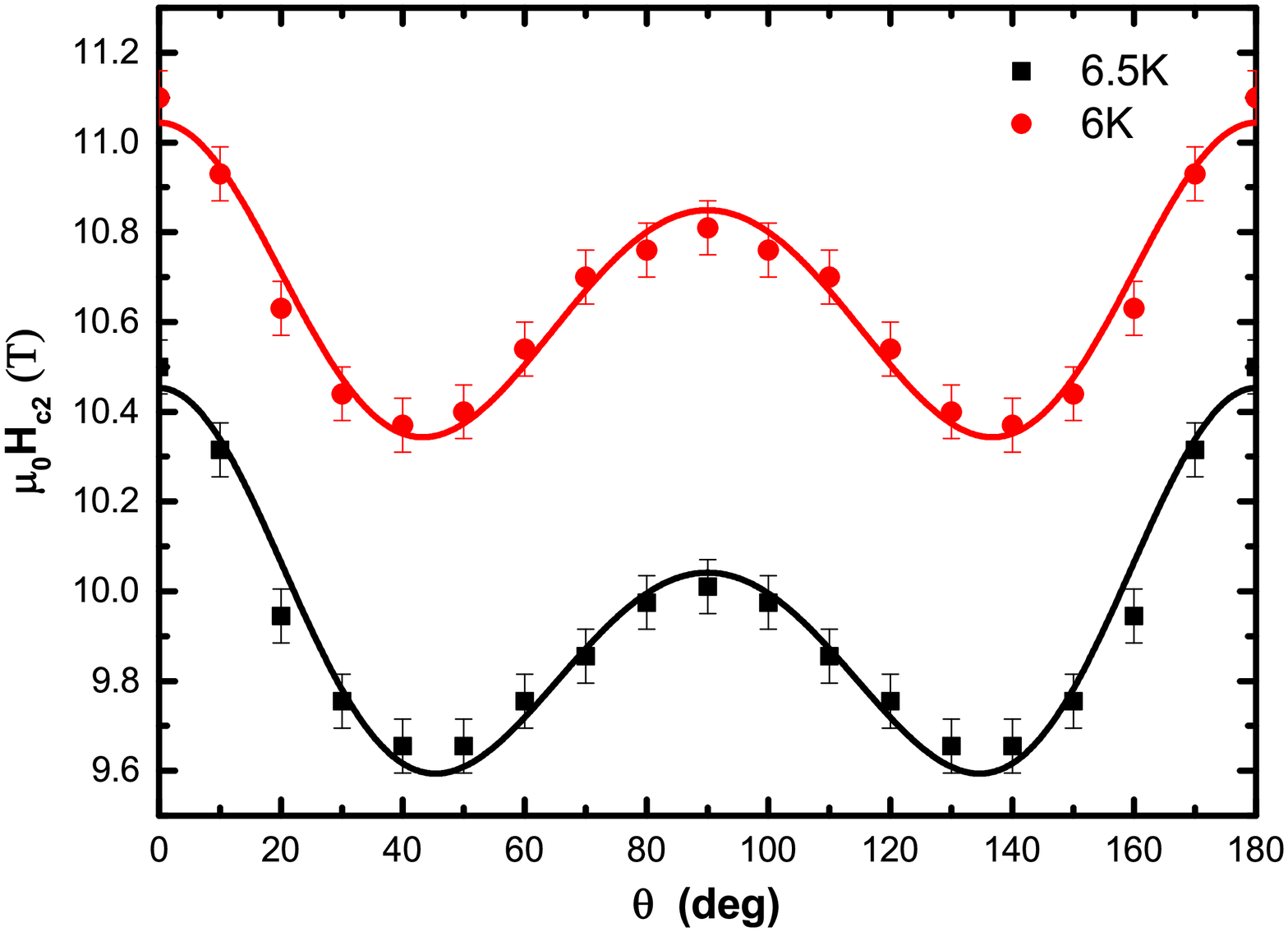}
\caption{Angular dependence of $H_{c2}$ at 6 and 6.5 K for RbCr$_3$As$_3$ single crystal B1. The solid lines are the fitted data using Eqs. (2) with $\mu_0H_{c2,\parallel}^{orb}$ = 13.8T, $\gamma$ = 0.74, $\mu_0H_{pm}^{\parallel}$ = 8.8 T for 6 K and $\mu_0H_{c2,\parallel}^{orb}$ = 13.7 T, $\gamma$ = 0.73, $\mu_0H_{pm}^{\parallel}$ = 8.8 T for 6.5 K, respectively.}
\label{fig:fig5}
\end{figure}

In order to check the angular dependence of $H_{c2}$ directly, field dependent magnetoresistance measurements were done with different angles between the magnetic field and the $c$ axis at 6 and 6.5K. The results of $H_{c2}(\theta)$ of  RbCr$_3$As$_3$ single crystal B1 are shown in Fig.\ \ref{fig:fig5}. $H_{c2}$ doesn't decrease monotonously as the field direction is tilted from $H \parallel c$ to $H \perp c$. Instead, a minimum appears between $\theta =$ 0 and $90^{o}$, which is similar to what has been observed in K$_2$Cr$_3$As$_3$.\cite{K2Cr3As3_zhu} To include the strong anisotropic paramagnetic pair-breaking effect, one can assume $H_{pm}(\theta)=H_{pm}^{\parallel}cos(\theta)$, where $H_{pm}$ is an effective Pauli-limiting field. Thus $H_{pm}=H_{pm}^{\parallel}$ for $\theta$ = 0, and $H_{pm}=0$ for $\theta = 90^{o}$. As $[H_{c2}(\theta)]^2=[H_{c2}^{orb}(\theta)]^2-[H_{pm}(\theta)]^2$,\cite{K2Cr3As3_zhu} one can get
\begin{multline}
H_{c2}(\theta) = \sqrt{\frac{(H_{c2,\parallel}^{orb})^{2}}{cos^{2}(\theta)+\gamma^{-2}sin^{2}(\theta)}-(H_{pm}^{\parallel}cos(\theta))^{2}}  \label{eq:Hc2theta}
\end{multline}
As shown in Fig.\ \ref{fig:fig5}, the $H_{c2}(\theta)$ data can be fitted very well by the above equation, confirming the absence of Pauli-limiting effect for $\theta = 90^{o}$. Similar results have been reported in K$_2$Cr$_3$As$_3$.\cite{K2Cr3As3_canfield, K2Cr3As3_zhu} Usually, the $H_{c2}(0)$ value is limited by
paramagnetic effect regardless of field directions for a conventional superconductor with a high $H_{c2}(0)$
comparable to $H_p$. Regarding the insensitivity of $T_c$ to nonmagnetic impurities and the behavior of $H_{c2}(T)$ in K$_2$Cr$_3$As$_3$, Balakirev \textit{et al.} proposed a novel spin-singlet superconductivity with electron-spin locking
along the $c$ direction.\cite{K2Cr3As3_canfield} Zuo \textit{et al.} pointed out that the spin state of $\mid\leftleftarrows>$ + $\mid\rightrightarrows>$ is equivalent to $\mid\uparrow\downarrow>$ + $\mid\downarrow\uparrow>$ with $S_z=0$ for the odd-parity Cooper pairs.\cite{K2Cr3As3_zhu} For $H \parallel c$, the Zeeman energy breaks the Cooper pairs, showing the Pauli-limiting behavior. However, for $H \perp c$, the field simply changes the population of Cooper
pairs with spin directions $\mid\leftleftarrows>$ and $\mid\rightrightarrows>$, and therefore, no
paramagnetic pair-breaking is expected.

\begin{table*}
\caption{\label{tab:table1}Summary of the parameters of RbCr$_{3}$As$_{3}$ and Rb$_{2}$Cr$_{3}$As$_{3}$. The data of Rb$_{2}$Cr$_{3}$As$_{3}$ are from Ref. 14.}
\begin{ruledtabular}
\begin{tabular}{c c c c c c c c c c c c c c c}
$ $  &$T_c$  &$\mu_0H_{c2}^{\parallel c'}|_{T_{c}}$  &$\mu_0H_{c2}^{\perp c'}|_{T_{c}}$ &$m_{\perp}/m_{\parallel}$ &$\xi_{\parallel c}(0)$ &$\xi_{\perp c}(0)$ &$\mu_0H_p$ &$\mu_0H_{c2}^{\parallel c}(0)$ &$\mu_0H_{c2}^{\perp c}(0)$  &$\gamma(0)$  &$\gamma(T_c)$ &$T(\gamma=1)$\\
       &(K)        &(T/K)        &(T/K)          &   &(nm)  &(nm)             &(T)                   &(T)           &(T)     &      &               &(K)\\
\hline
RbCr$_{3}$As$_{3}$          &7.3       &-18        &-8.5     &4.5  &3.9    &1.9                 &13.4            &27.2         &43.4      &1.6       &0.5                &0.75$T_c$\\
\\
Rb$_{2}$Cr$_{3}$As$_{3}$    &4.8       &-16        &-3       &28  &3.2    &2.1                  &8.8            &17.5           &29      &1.7       &0.19                &0.4$T_c$ \\

\end{tabular}
\end{ruledtabular}
\end{table*}

%We noted that $H_{c2}$  shows a three-fold modulation as a function of the azimuthal angle $\phi$ in K$_2$Cr$_3$As$_3$.\cite{K2Cr3As3_zhu} Thus we also did similar measurements at 6 and 6.5K with magnetic field up to 14T.  However, we didin't see any three-fold or six-fold modulation within the measurement resolution. Perhaps the modulation is more obvious at very low temperature.  Unfortunately, the magnetic field we can use now is not high enough to do this measurement at very low temperature.

Until now, different paring mechanism and symmetry has been proposed for A$_2$Cr$_3$As$_3$ compounds, such as p$_{z}$-wave spin-triplet,\cite{Molecular,zhou,p-wave} spin singlet\cite{Stongcoupling-K2Cr3As3, K2Cr3As3_canfield} and a two-band model.\cite{two-band,wen}. The experimental results have not yet reached a consensus.\cite{review,NLWang}
Extremely air sensitivity property of A$_2$Cr$_3$As$_3$ hinders many further studies for their intrinsic physical characteristics.
In Table I, all the parameters of RbCr$_{3}$As$_{3}$ we have obtained are summarized and compared to Rb$_{2}$Cr$_{3}$As$_{3}$.
Most of the parameters of the two compounds are close except $m_{\perp}/m_{\parallel}$, especially the ratio of $H_{c2}(0)/T_c$ (for both $H \parallel c$ and $H \perp c$) are almost the same.
Although ACr$_3$As$_3$ has a centrosymmetric crystal structure differing
from its noncentrosymmetric counterpart A$_2$Cr$_3$As$_3$, all of the results we obtained above indicate that the superconducting property of RbCr$_3$As$_3$ is very similar to the Rb$_2$Cr$_3$As$_3$ compounds.
Investigations on the air stable ACr$_3$As$_3$ compound may provide a good path to acquire deep insight
into the superconducting mechanism in the
Q1D Cr-based family, and may help to expand the overall understanding of unconventional superconductivity.

\section{Summary}

In summary, we have constructed the $H_{c2}-T$ phase diagram for RbCr$_{3}$As$_{3}$
with $T_c \approx 7.3$ K by use of magnetoresistance measurements with temperature down to 0.35 K in
static magnetic fields up to 38 T both for directions parallel and perpendicular to the
$c$ axis. Fitting with the WHH model
yields zero-temperature critical fields of $\mu_{0}H_{c2}^{\parallel c}(0)\approx $ 27.2 T
and $\mu_{0}H_{c2}^{\perp c}(0)\approx $ 43.4 T, both exceeding the BCS weak-coupling
Pauli limit. The anisotropy of $H_{c2}$ has a reversal at $\sim $ 5.5 K.  The paramagnetic pair breaking effect is strong for $H \parallel c$ but absent for $H \perp c$, which was further confirmed by the $H_{c2}(\theta)$ data.

This work was supported by the National Science Foundation
of China (Nos. 11704385, 11874359 and 11774402). A portion of this work was performed on the Steady High Magnetic Field Facilities, High Magnetic Field Laboratory, Chinese Academy of Sciences, and supported by the High Magnetic Field Laboratory of Anhui Province.

\end{document}